\documentstyle[prl,aps,epsf,multicol,tabularx,psfig,graphicx]{revtex} 
\tighten 
\begin{document} 
\draft 
\title{Phase Slips and Interfaces in $D$--Wave Superconductors.} 
 
\author{Andrew M. Martin$^{\dagger}$ and James F. Annett$^{\S}$} 
\address{$^{\dagger}$D\'{e}partement de Physique Th\'{e}orique, 
Universit\'{e} de Gen\`{e}ve, 1211 Gen\`{e}ve 4, Switzerland.} 
\address{$^{\S}$University of Bristol, H.H. Wills Physics Laboratory, 
Royal Fort, Tyndall Ave, 
Bristol BS8 1TL, United Kingdom.}

\date{\today} 
\maketitle 
 
\begin{abstract} 
We consider a model (100) interface between two $d$--wave 
superconductors. By solving the Bogoliubov de Gennes equation on a 
tight binding lattice, we study the properties of the interface as 
a function of the interface barrier. We contrast the two 
scenarios: (i) an order parameter phase difference of $\theta=0$ 
across the interface, and (ii) a phase slip of $\theta=\pi$ across 
the interface. We find resonant sub-gap structure in the density 
of states only when there is a phase slip present. We show that 
the local $s$--wave and ``$p$--wave like'' order parameter 
components are strongly influenced by the barrier profile and by 
the phase slip. The temperature dependence of the local $p$--wave 
order parameter  follows the underlying bulk $T_c$ for $p$--wave 
superconductivity implying that this could be measured by a 
suitable tunneling experiment. We also calculate the Josephson 
critical current as a function of the strength of the insulating 
barrier at the interface. 
\end{abstract} 
\pacs{Pacs numbers: 74.50.+r, 74.60.Jg, 74.80.-g} 
 
\begin{multicols}{2} 
\narrowtext 
\section{Introduction.} 
\label{introduction} 
\setcounter{footnote}{1} 
 
Over the past few years 
it has been generally 
agreed that the macroscopic symmetry of the superconducting order parameter 
 in high ${\rm T_c}$ superconductors 
is unconventional. The pairing state is almost certainly $d$--wave 
\cite{Scal95,Harl95,jfa96}, either pure $d_{x^2-y^2}$ or a mixed 
state such as $d+s$\cite{kouznetsov}  or $d+\imath 
s$\cite{goldman}, which is predominantly $d$--wave. The $d$--wave 
pairing state raises many important questions, such as the 
relationship between the pairing state and the physical properties 
of the superconductors, and of course, the microscopic origin of 
the pairing interaction itself. 
 
In particular it becomes interesting to consider the microscopic 
properties of the interfaces of $d$--wave symmetry systems.  This 
is necessary both to understand the physical properties of high 
$T_c$ materials, and to develop further tests of the pairing state 
and order parameter. For example, it has already been shown that 
at an normal metal to $d$--wave superconductor ($N-D$) interface 
there is an extended $s$--wave component to the order parameter in 
the region of the interface and that this survives up to a few 
coherence lengths away from the interface \cite{MS1,MS2,MS3,us}. 
There is also experimental and theoretical evidence for 
microscopic time reversal symmetry breaking at such 
interfaces\cite{covington,BBS}. The presence or absence of such 
subdominant order parameters can, in principle, be used to extract 
information about the nature of the microscopic pairing 
interaction. For example some models will allow both $d$ and 
$s$--wave pairing, while other models are pair breaking in the 
$s$--wave channel. There has also been a substantial amount of 
work concerning the effect of $d$--wave symmetry on the Josephson 
Effect \cite{YTanaka,BGZ,ZWT,TWZ,Huck97}, mesoscopic scattering 
properties of $N-I-D$ structures \cite{CRHu,XMT,BSB,Purdue} and 
tunneling  density of states for surfaces of $d$--wave 
superconductors \cite{Tanuma1,Tanuma2}. 
 
In this paper we examine 
the self-consistent changes in 
\begin{figure} 
\includegraphics[scale=1.6]{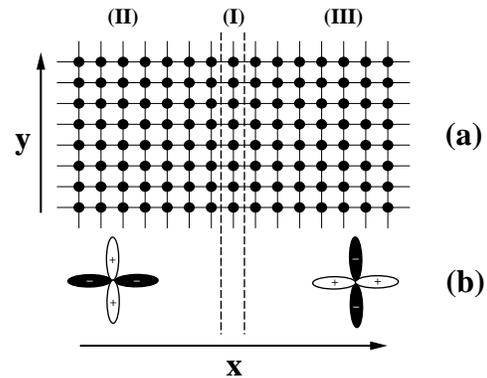} 
\caption{A schematic diagram of the system we wish to consider. Regions I, II 
and III represent a tight binding lattice, where regions II and III are 
connected to each other via an interface (region I). In regions II and III we 
can define the symmetry of the superconducting order parameter to be 
mismatched by $\pi$.\label{figone}} 
\end{figure} 
\noindent the superconducting order parameter near an 
insulating barrier between two $d$--wave superconductors, i.e. a 
$D-I-D$ interface.  The basic geometry is shown in Fig. 1. We 
carry out a detailed study of how the interface properties evolve 
as the strength of the barrier is increased from a small 
perturbation to a strong tunneling barrier. In this way we can 
contrast the behaviour of interfaces with little or no tunnel 
barrier, such as twin boundaries, to interfaces with a strong 
tunnel barrier, such as tunneling through vacuum or oxygen 
depleted regions at the surface. Our results may also be relevant 
as a model of grain boundary junctions in high $T_c$ materials, 
since Gurevich and Pashitskii have proposed that these junctions 
are essentially $D-I-D$ junctions whose impurity barrier is simply 
dependent up the angle of the grain boundary\cite{interface}. 
 
In our calculations we will concern ourselves with two scenarios 
(i) where there is no phase difference in the order parameter over 
the interface ($\theta=0$) and (ii) where there is an imposed 
phase slip of $\pi$ over the interface ($\theta=\pi$), as 
illustrated in Fig.~\ref{figone}. For both of these scenarios we 
focus our attention on the magnitude of the extended $s$--wave and 
``$p$--wave  like'' components of the order parameter in the 
region of the interface. There is already a great interest in the 
role extended $s$--wave components have to play in the region of 
twin-boundaries, for example W. Belzig, C. Bruder and M. Sigrist 
\cite{BBS} have studied how the local density of states is 
modified, as one moves across a twin boundary when the bulk order 
parameter is $d \pm \imath s$ and the effect of temperature on 
these calculations. 
 
The plan of this paper is as follows. In Sec. II we describe briefly 
the model Hamiltonian, and the methodology used to 
solve the Bogoliubov de Gennes equation in our system. 
Then in Sec. III we will describe, in detail, the system of interest and 
define the different components of the 
superconducting order parameter which will be relevant. In Sec. IV 
we present results for a perfect interface, i.e. the limit 
that the tunnel barrier height is zero. 
We note that in this 
{\it clean} case 
an order parameter phase change of 
 $\theta=\pi$ across the ``interface'' corresponds to a 
phase slip of $\pi$ in a perfect superconductor. We establish that 
this phase slip is a stable soliton like solution of the 
Bogoliubov de Gennes equation and we obtain the local density of 
electronic states in the vicinity of the phase slip. We then 
proceed to analyze the different components to the superconducting 
order parameter at the centre of the phase slip and how these 
change as we change the temperature of the system. In Sec V we 
turn our attention to $D-I-D$ interfaces, calculating the 
properties of the interface as a function of the tunnel barrier 
height for both the scenarios $\theta=0$ and $\theta=\pi$. Finally 
we conclude with comparisons of the local particle density of 
states in the region of the interface. Using these local particle 
density of states we also estimate how the Josephson critical 
current is changed as the tunnel barrier strength is changed. 
 
\section{The Bogoliubov de Gennes Equation.} 
\label{bdge}

Our starting point is the 
Bogoliubov de Gennes (BdG) 
equation, 
\begin{equation} 
\sum_{j} {\bbox H_{ij}} 
\left( \matrix{ 
u_{j}^{n} \cr 
v_{j}^{n}} 
\right) 
= 
E_{n} 
\left( 
\matrix{ 
u_{i}^{n} \cr 
v_{i}^{n}} 
\right) \label{eq:1} 
\end{equation} 
where $u_i^n$ and $v_i^n$ are the particle and hole amplitudes on 
site $i$ associated with eigenenergy $E_n$. 
The Hamiltonian is 
\begin{equation} 
\label{eq:1a} 
{\bbox H_{ij}}= \left( \matrix{ 
H_{ij} & \Delta_{ij}  \cr 
\Delta^{\star}_{ij} & -H^{\star}_{ij}} \right), 
\end{equation} 
where $H_{ij}$ is the normal part of the tight binding Hamiltonian 
and $\Delta_{ij}$ is the anomalous part, given by the 
BCS gap function.

In the case of the 
cuprates we have 
$d_{x^2-y^2}$ pairing on a square two-dimensional lattice. 
The simplest model Hamiltonian which leads to this pairing state 
is the non-local attractive Hubbard model, 
\begin{equation} 
 \hat{H} = \sum_i \epsilon_i 
 c^\dagger_{i\sigma} c_{i\sigma} 
+ \sum_{<ij>} \left( t_{ij}( 
 c^\dagger_{i\sigma} c_{j\sigma} 
+ {\rm h.c.}) 
       -\frac{U_{ij}}{2} \hat{n}_i\hat{n}_j \right), 
\label{eq:hubbard} 
\end{equation} 
where, as usual, $c^\dagger_{i\sigma}$ and $c_{i\sigma}$ are the 
electron creation and annihilation operators at site $i$ for spin 
$\sigma$ and $\hat{n}_i$ is the number operator at site $i$. For 
simplicity we shall assume that both $t_{ij}$ and $U_{ij}$ are 
limited to nearest neighbors only. We shall neglect any on-site 
interactions $U_{ii}$. Also, for simplicity, we neglect 
retardation in the attractive interaction.

The BdG equation, Eq.~\ref{eq:1} is easily derived as the appropriate 
mean field factorization of the Hubbard model Eq.~\ref{eq:hubbard}. 
The normal 
 part of the Hamiltonian Eq.~\ref{eq:1a}, 
$H_{ij}$, is given by 
\begin{equation} 
\label{eq:2} 
H_{ij}= 
(\epsilon_{i}' - \mu) \delta_{ij} 
+t_{ij}'(1-\delta_{ij}) 
\end{equation} 
where $\mu$ is the 
chemical potential. 
 The on-site energy, $\epsilon_i'$, 
includes the bare on-site energy and the normal Hartree potential 
\begin{equation} 
   \epsilon_i' = \epsilon_i + \sum_j U_{ij} n_{jj} 
\end{equation} 
where $n_{jj}$ is the charge density at site $j$. 
Similarly the self-consistent hopping $t_{ij}'$ is 
the bare hopping plus a Hatree-Fock exchange term 
\begin{equation} 
t_{ij}'=t_{ij}+\frac{1}{2} U_{ij} n_{ij}. 
\end{equation} 
The charge densities $n_{ii}$ and  $n_{ij}$ are determined from the 
eigenvectors of the Hamiltonian Eq.~\ref{eq:1} and are 
given by 
\begin{eqnarray} 
n_{ij} &=& \sum_{\sigma}\langle\Psi_{i \sigma}^{\dagger}\Psi_{j \sigma}\rangle 
\nonumber \\ 
&=& 
2\sum_{n}[(u_i^n)^{\star}u^n_j f(E_n)+v_i^n(v_j^n)^{\star}(1-f(E_n))] 
\label{eq:3} 
\end{eqnarray} 
where the sum is over all of the {\em negative} eigenvalues 
$E_n$ only. 
 
The off-diagonal part of the mean field Hamiltonian 
Eq.~\ref{eq:1a} is the pairing potential or gap function, 
$\Delta_{ij}$. 
%
Self-consistently this is determined 
from 
\begin{equation} 
\label{eq:4} 
\Delta_{ij}=-U_{ij}F_{ij} 
\end{equation} 
where the anomalous density is 
\begin{eqnarray} 
F_{ij} &=& \langle\Psi_{i \uparrow}\Psi_{j \downarrow}\rangle \nonumber \\ 
&=& 
\sum_{n}\left[u_i^n (v^n_j)^{\star}(1- f(E_n)) 
-(v_i^n)^{\star}(u_j^n)f(E_n)\right] 
\label{eq:5} 
\end{eqnarray} 
and again the sum only includes 
eigenvalues $E_n$ up to the condensate chemical potential ($\mu$).

In the case of a bulk square lattice with nearest neighbour 
attractions $U_{ij}$ and nearest neighbour hopping the solutions 
to the above set of self-consistent equations are well known. 
Depending on the band filling and temperature the gap function 
$\Delta_{ij}$ may be pure $d_{x^2-y^2}$, pure extended $s$--wave 
or a $d+\imath s$ mixed state \cite{Jon1}. In this paper we shall 
work exclusively near half filling where only the pure $d$--wave 
state is stable.

To solve the above system of self-consistent equations we employ the 
Recursion Method \cite{us,RHaydock,annett,LMG} 
following exactly the method outlined in \cite{us}. 
The advantage of this method is that it is a purely real-space calculation, 
and so it is not necessary to assume periodic boundary conditions. 
For example, in the interface geometry studied in this paper 
we have 
two semi-infinite lattices joined at an interface. 
The recursion method also gives directly the local density of states, $N(E)$, 
at any given site, which, as we shall show below, assists 
in the physical interpretation of the self-consistent numerical results. 
 
 
 
\section{Interface geometry.} 
 
In the cuprate superconductors there are a number of important 
intrinsic weak links, including twin boundaries, low angle grain 
boundaries, and edge dislocations. There are also a number of man 
made interfaces designed for specific applications, including 
break junctions, high angle grain boundary junctions, ramp 
junctions, and junctions with electron and ion beam irradiated 
barrier regions. In separate papers we have investigate simple 
$S-N$, $D-N$ and $S-D$ interfaces\cite{us}, and examined in detail 
the behaviour of a $D-D$ high angle grain boundary 
junction\cite{jason}. In this paper we seek to examine how the 
interface properties are modified by the presence of an insulating 
barrier. For example the insulating barrier may be caused by 
oxygen depletion in the region of the interface, as suggested by 
Gurevich and Pashitskii in their model of grain 
boundaries\cite{interface}.

With this motivation, we shall study in detail the system shown in 
Fig.~\ref{figone}(a). We have two semi-infinite $d$--wave 
superconducting regions (II) and (III) which are connected to each 
other via an interface (I). We study (100) oriented interfaces as 
shown in Fig.~\ref{figone}(a). In the interface region (I) we 
define the strength of the barrier, $V$, 
 by setting 
$\epsilon_i'=V$ in this region to be finite, compared to the rest 
of the system where $\epsilon_i'=0$. This barrier height could 
physically represent the amount of oxygen depletion at the 
interface, or the vacuum or insulating barrier width in the case 
of (100) surface tunnel junctions.  We shall examine the changes 
in the interface properties as the barrier height $V$ is varied 
from zero to a large value. 
 
We shall study the Josephson energy of the junction by 
comparing the cases of either zero or $\pi$ 
phase difference between the bulk order parameters 
far from the interface.  Fig.~\ref{figone}(b) illustrates the case 
of a $\pi$ phase difference in the $d$--wave order parameter. 
We carry out the calculation by imposing a fixed bulk $d$--wave 
order parameter $\Delta_{ij}$ far from the interface, and then 
self-consistently computing the $\Delta_{ij}$ in the region 
near to the interface. Typically we perform the self-consistent calculations 
in a region of width five coherence lengths on either side 
of the junction, confirming that after this distance the 
self-consistent parameters have returned to their bulk values. 
 
The self-consistent order parameters $\Delta_{ij}$ are computed for 
each nearest neighbour bond $<ij>$ in the lattice 
($\Delta_{ij}=\Delta_{ji}$). However interpreting these results 
is simplified if we decompose these $\Delta_{ij}$ into components 
of different symmetries. For this reason we define the following 
four linear combinations at each site $i$, 
\begin{eqnarray} 
\Delta^{(d)}_i &=& 
\frac{1}{4}\Delta_{ij_{1}}-\Delta_{ij_{2}}+\Delta_{ij_{3}}- 
\Delta_{ij_{4}} \nonumber \\ 
\Delta^{(s)}_i &=& \frac{1}{4}\Delta_{ij_{1}}+\Delta_{ij_{2}}+ 
\Delta_{ij_{3}}+\Delta_{ij_{4}} \nonumber \\ 
\Delta^{(p_x)}_i& =& \frac{1}{2}\Delta_{ij_{1}}-\Delta_{ij_{3}}\nonumber \\ 
\Delta^{(p_y)}_i&=&\frac{1}{2}\Delta_{ij_{2}}-\Delta_{ij_{4}}. 
\nonumber 
\end{eqnarray} 
Here the sites $j_1 \dots j_4$ are the four nearest neighbours 
of site $i$, say counting clockwise. 
The first two  combinations are simply $d_{x^2-y^2}$ 
and extended $s$--wave pairing respectively. The second two 
correspond to components of the $\Delta_{ij}$ which are {\em odd} 
about site $i$. For this reason they may be said to be 
``$p$--wave {\it like}'' components of the order parameter. 
It is important to note that 
these components do not 
 truly correspond to $p$--wave pairing since they are spin singlet 
not triplet and there is no spontaneous breaking of spin 
rotational symmetry. These two terms are more closely related to 
{\em gradients} of $s$ or $d$--wave order parameters. They also 
have some relationship to Yang's $\eta$ pairing order 
parameter\cite{yang} which is a spin singlet order parameter, but 
one that alternates sign on neighboring lattice sites. This has 
also been referred to elsewhere as {\em 
antiferrosuperconductivity}\cite{gyorffy,annett}. Here 
$\Delta^{(p_x)}$ and $\Delta^{(p_y)}$ correspond to an alternating 
non-local order parameter $\Delta_{ij}$, unlike $\eta$-pairing 
which corresponds to  a staggered $\Delta_{ii}$ on-site gap 
function. Note that because the system in Fig.~\ref{figone}(a) is 
perfectly periodic in $y$ we shall always have 
$\Delta^{(p_y)}_i=0$. 
 
For our chosen model parameters only the $d$--wave component 
$\Delta^{(d)}_i$ is non-zero in the bulk far from the interface. However 
components of the other types can arise near the interface 
driven by the local symmetry breaking. The self-consistent 
order parameter can also have spontaneous symmetry breaking phase 
transitions as a function of temperature or chemical potential, 
for example breaking of time reversal symmetry into an $d+\imath s$ type 
state.

\section{Interface Properties: No Barrier} 
\label{sec3} 
 
Having outlined in previous sections the method for calculating 
and analyzing the properties of the order parameter, we now 
consider the most simple situation: $V=0$ throughout the system. 
In this case there is no barrier at all at the interface $I$. If 
the bulk phase difference across the system $\theta$ is zero then 
we obviously have just a simple bulk system everywhere with the 
bulk $d$--wave order parameter. 
 
On the other hand if $\theta=\pi$ (Fig.~\ref{figone}(b)) then we 
have a {\em phase slip} in the unit cell. This is the 
superconducting analogue of a Bloch wall separating magnetic 
domains. By solving the Bogoliubov de Gennes equations 
self-consistently we can calculate the width of the phase slip 
region and the variation of the order parameter across the phase 
slip. 
 
\begin{figure} 
\vspace{-2.0cm} 
\includegraphics[scale=0.35,angle=-90]{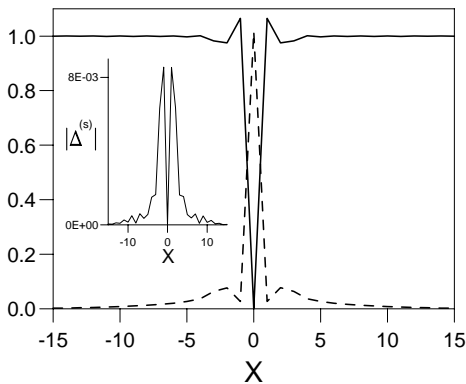} 
\caption{Profiles of different symmetry components of the 
superconducting order parameter in the presence of a 
$\theta=\pi$ phase slip, 
$|\Delta^{(d)}(x)|/|\Delta^{(d)}(\infty)|$ (solid line) and 
$|\Delta^{(p_x)}(x)|/|\Delta^{(d)}(\infty)|$ (dashed line). 
The insert shows $|\Delta^{(s))}(x)|$ in the region of the interface. 
For all of these figures $T=0$ and the barrier height is $V=0$.\label{figtwo}} 
\end{figure}

Fig.~\ref{figtwo} shows our self-consistently determined order parameters 
in this phase slip geometry. 
Since all quantities only vary 
in the $x$ direction, we can plot $|\Delta^{(d)}(x)|$, 
$|\Delta^{(s)}(x)|$ 
$|\Delta^{(p_x)}(x)|$ in the region of the interface ($x=0$), noting that 
$|\Delta^{(p_y)}(x)|=0$. 
We started the self-consistent calculation by imposing a $d$--wave 
order parameter throughout the system, but with a phase change of 
$\pi$ at $x=0$. The solution was then determined self-consistently 
by iterating the BdG equations starting from this state. We find 
that the phase slip is a stable self-consistent solution of the 
BdG equations. 
 
In Fig.~\ref{figtwo} we can see how the $|\Delta^{(d)}(x)|$ (solid line) 
varies in the region of the interface. 
It is clearly a minimum  at the interface and quickly reaches it's bulk value, 
oscillat-
\noindent ing as it does \cite{OJ,YM}. The extended $s$-wave component 
 $|\Delta^{(s)}(x)|$ (insert) is also a 
minimum at the interface, but it then reaches a maximum close to the interface 
and then reduces to zero in the bulk. On the other hand 
$|\Delta^{(p_x)}(x)|$ (dashed line) 
has a strong  maximum at the interface. 
\begin{figure} 
\vspace{-2.0cm} 
\includegraphics[scale=0.35,angle=-90]{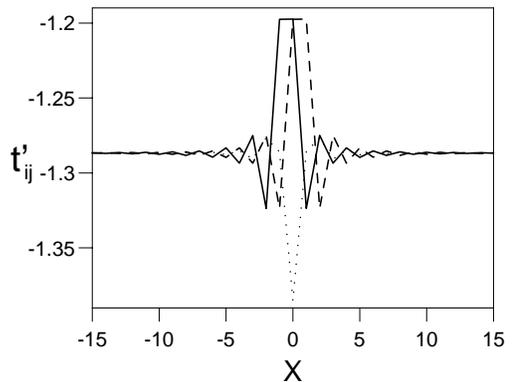} 
\caption{The effective hopping $t^{\prime}_j$ as a function of distance from 
the interface ($x=0$). The hoppings are 
$t^{\prime}_{j1} (x)$ (solid line), $t^{\prime}_{j2}(x)$ (dotted line), 
$t^{\prime}_{j3}(x)$ (dashed line) and $t^{\prime}_{j4}(x)$ (dotted line), 
where $j_1$ to $j_4$ correspond to the four different 
nearest neighbour bonds. All parameters are the same as in Fig. 2. 
\label{figthree} } 
\end{figure} 
 \vspace{-0.5cm}
 
It is also interesting to consider the self-consistent hopping, 
$t^{\prime}_{ij}$, in the region of the barrier. 
Again, considering that each site $i$ has four neighbours 
$j_1 \dots j_4$ we can plot the four components 
$t^{\prime}_{ij_1} \dots 
t^{\prime}_{ij_4}$ as a function of position. Changing notation 
slightly we can think of these as four functions of position, $x$, 
$t^{\prime}_{j}(x)$ ($j=1..4$). 
In Fig. 3  the 
dashed and solid lines show $t^{\prime}_{1}(x)$ and $t^{\prime}_{3}(x)$ 
respectively (the hopping perpendicular to the interface) and the dotted line 
is $t^{\prime}_{2}(x)$ ($= t^{\prime}_{4}(x)$), 
 (the hopping parallel to the 
interface). The plot shows that in the region of the interface the 
hopping is modified from its bulk value. It is this modification 
of the hopping which helps stabilize the phase slip, this is 
analogous to the case of conventional local  $s$--wave pairing 
where modification of the local density in the region of the phase slip 
enhances its stability \cite{needreference}. 
The oscillations in the 
$t^\prime_{ij}$ shown in Fig. 3 simply correspond to Friedel 
oscillations in the non-local Hartree-Fock exchange term 
($\frac{U_{ij}n_{ij}}{2}$). 
 These oscillations 
are exactly analogous to the Friedel oscillations around an impurity 
or surface in a normal 
system, even though there is no {\em normal} barrier $V$ here. 
 
The results shown in Figs.~\ref{figtwo} and \ref{figthree} were performed 
with $T \ll T_c^d$, the bulk $d$--wave transition temperature. 
 We now proceed to examine the temperature dependence. 
 In Fig. 4 we have plotted the normalized 
temperature dependence of $|\Delta^{(p_x)}(0)|^2$ (circles) and 
$|\Delta^{(d)}(\infty)|^2$ (crosses), these being the $p$--wave 
{\it like} order 
parameter at the interface and the bulk $d$--wave order parameter 
respectively. As we can see 
$|\Delta^{(p_x)}(0)|^2$  is a linear function of temperature 
over a wide range (from $T\sim0.1t$ to $0.4t$). Extrapolating this linear 
dependence we can state that  $|\Delta^{(p_x)}(0)|^2$  would go to zero at a temperature of 
around
\begin{figure} 
\vspace{-3.0cm} 
\includegraphics[scale=0.35,angle=-90]{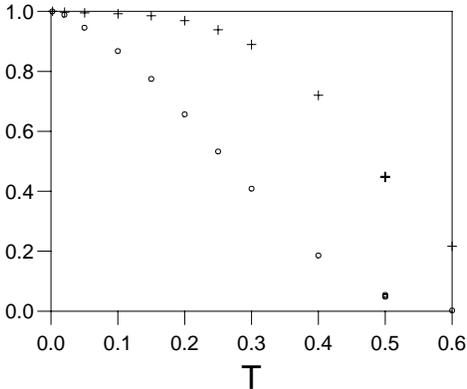} 
\caption{The temperature dependence of  the $p$--wave component of 
the order parameter, 
$\frac{|\Delta^{(p_x)}(0)|^2}{|\Delta^{(p_x)}(0)|^2_{T=0}}$ 
(circles), and the $d$--wave component, 
$\frac{|\Delta^{(d)}(\infty)|^2}{|\Delta^{(d)}(\infty)|^2_{T=0}}$ 
(crosses), $T$ is measured in units of $t$. \label{figfour}} 
\end{figure} 
\noindent   $T_c^p \approx 0.5t$, if it were not coupled to the 
$d$--wave order parameter. 
From Fig. 4 we  infer 
from the fact that $|\Delta^{(p_x)}(0)|^2$ deviates from a 
straight line above $T\approx 0.5t$ that it is enhanced 
above $T_c^p$ by coupling to the $d$--wave order parameter.

This unusual temperature dependence of the $p$--wave like 
component can be explained by examining the phase diagram for a 
perfect square tight binding lattice. It is known that at half 
filling we have a ground state with a stable $d$--wave order 
parameter. However one can also define a separate $T_c$ for 
extended $s$--wave or $p$--wave states. Near half filling $T_c^d$ 
is greater than $T_c^s$ or $T_c^p$ and so only the $d$--wave state 
occurs in the bulk\cite{Jon1}. However if the $d$--wave and 
extended $s$--wave order parameters were suppressed somehow then 
the system would in principle go superconducting at $T_c^p$. 
Interestingly the $T_c$ for the ``$p$--wave like'' singlet order 
parameter $\Delta^{(p)}$ is the same as that of true triplet 
$p$--wave order parameter, which follows from the identical form 
of the gap equation in both cases\cite{Jon1}. In effect our phase 
slip has forced $\Delta^{(d)}$ and $\Delta^{(s)}$ to vanish at 
$x=0$. This allows the intrinsic underlying $p$--wave state to 
become apparent. It would be interesting to test this underlying 
$p$--wave state experimentally, for example by tunneling 
experiments to see whether or not there is a finite gap inside a 
superconducting phase slip. If there is such a gap it would show 
that the pairing mechanism has attractive components for $p$--wave 
pairing as well as $d$--wave. This would only be true for some 
specific model pairing interactions, such as our nearest neighbour 
Hubbard model.

\section{Interface Properties: Dependence on Barrier.}

Having considered the most simple case ($\epsilon_i=0$ for all 
sites) we now focus our attention on what happens if we have a 
finite tunnel barrier ($V \ne 0$) in region (I) of our system 
(Fig.~\ref{figone}). In this case it is also interesting to 
consider the situation where there is no phase slip over the 
interface and compare this to the result obtained when there is a 
phase slip. In this section we study three different strengths of 
barrier, where in region (I) we have defined $V/t=1,2,5$. 
\begin{figure} 
\includegraphics[scale=0.35,angle=-90]{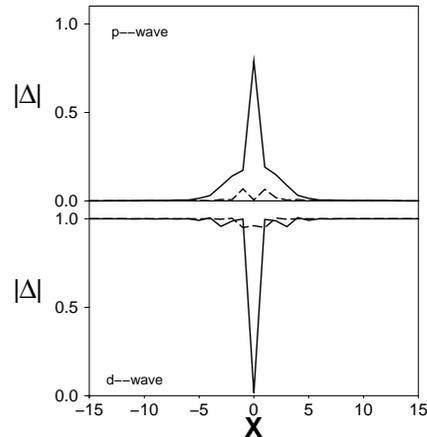} 
\caption{Profiles of different symmetry components of the 
superconducting order parameter in the presence of no 
phase slip, $\theta=0$, (solid lines), and a 
$\theta=\pi$ phase slip (dashed lines). 
The upper graph plots 
$|\Delta^{(p_x)}(x)|/|\Delta^{(d)}(\infty)|$ and the lower graph plots 
$|\Delta^{(d)}(x)|/|\Delta^{(d)}(\infty)|$. 
The barrier interface height is $V=t$ and temperature is 
$T=0$. 
\label{figfive}} 
\end{figure}
In Fig.~\ref{figfive} we compare  both $|\Delta^{(p_x)}(x)|$ and 
$|\Delta^{(d)}(x)|$ for the cases of (i) there is a phase slip 
across the interface (solid line) and (ii) when there is no phase 
slip (dashed line), for $V/t=1$. From this figure we can see that 
as in the case of no barrier the phase slip means that 
$|\Delta^{(d)}(0)|=0$. This then, effectively, induces a finite 
$|\Delta^{(p_x)}(0)|$ at the interface. However when there is no 
phase slip (dashed lines) we can see that although the 
$|\Delta^{(d)}(x)|$ is diminished in the region of the interface 
it does not go to zero, conversely now, due to the symmetry of the 
problem, $|\Delta^{(p_x)}(0)|$ must be zero. We have also 
calculated the extended $s$--wave component to the order parameter 
in the region of the interface. We find that  the extended 
$s$--wave component is a maximum, at the interface, when there is 
no phase slip and zero when there is a phase slip, similar to 
Fig.~\ref{figtwo}. 
 
\begin{figure} 
\includegraphics[scale=0.35,angle=-90]{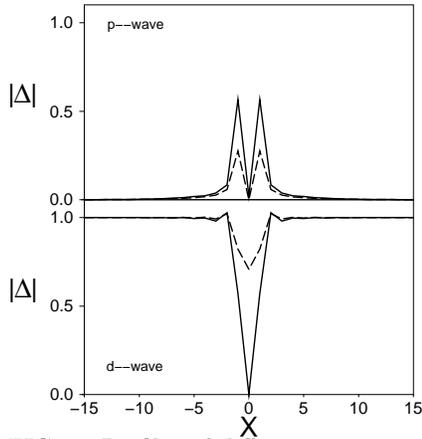} 
\caption{Profiles of different symmetry components of the 
superconducting order parameter, for an intermediate barrier height, 
 $V=2t$.  The plotted curves are otherwise exactly as in Fig. 5. 
\label{figsix}} 
\end{figure}

\begin{figure}  
\includegraphics[scale=0.35,angle=-90]{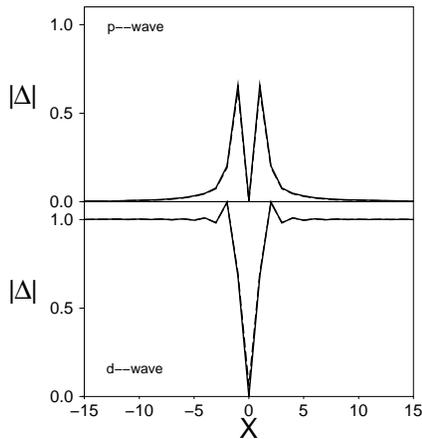} 
\caption{Profiles of different symmetry components of the 
superconducting order parameter, for a large barrier height, 
 $V=5t$.  The plotted curves are otherwise exactly as in Fig. 5. 
\label{figseven}} 
\end{figure}

Figs.~\ref{figsix} and \ref{figseven} show the different 
components to the order parameter for larger strengths of barrier 
($V/t=2,5$). What we see in these two figures is that as the 
barrier height is increased the differences between the $\theta=0$ 
and $\theta=\pi$ cases becomes smaller. In the limit of a very 
high strength impurity barrier, Fig.~\ref{figseven}, there is 
essentially no difference between the order parameter components 
 whether a phase slip is 
present or not. Fig.~\ref{figsix}, however, is more interesting. We see that 
 when there is 
no phase slip  $|\Delta^{(d)}(x)|$ is reduced in the region of the 
barrier, and as expected $|\Delta^{(p_x)}(0)|$  is zero. In the 
presence of a phase 
\noindent slip, as in all the other cases, the 
$|\Delta^{(d)}(0)|$ is zero at the phase slip. But, unlike 
Figs.~\ref{figtwo} and ~\ref{figfive}, $|\Delta^{(p_x)}(0)|$ is 
also zero. This shows that we have gone through three regimes at 
the interface.  These being (i), low scattering at the interface 
where, $|\Delta^{(d)}(0)|$ is finite when there is no phase slip 
and $|\Delta^{(p_x)}(0)|$ is finite in the presence of a phase 
slip, (ii), intermediate scattering at the interface where, 
$|\Delta^{(d)}(0)|$ is still finite when there is no phase slip 
but now $|\Delta^{(p_x)}(0)|$ is zero in the presence of a phase 
slip and (iii), large scattering at the interface, 
$|\Delta^{(d)}(0)|$ is now zero when there is no phase slip and 
$|\Delta^{(p_x)}(0)|$ is also zero in the presence of a phase 
slip. 
 
\begin{figure}
\includegraphics[scale=0.35,angle=-90]{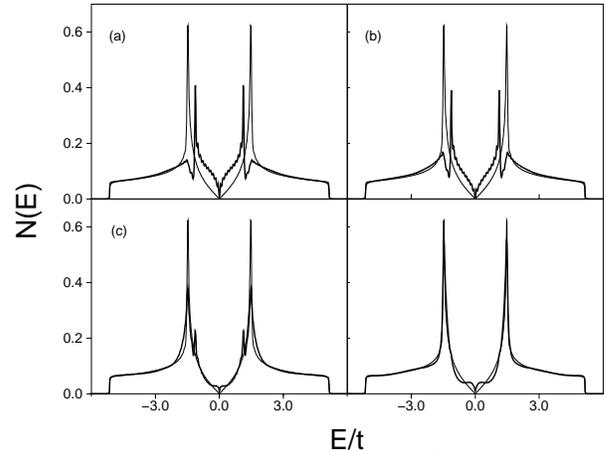} 
\caption{Local particle density of states in the region of the interface 
when the barrier height is $V=0$. We consider both  a phase 
slip (thick solid line) and no phase slip (thin solid line). The curves 
plotted are (a) 
$x=0$, (b) $x=1$, (c) $x=2$ and (d) $x=3$.\label{figeight}} 
\end{figure}

\begin{figure}
\includegraphics[scale=0.35,angle=-90]{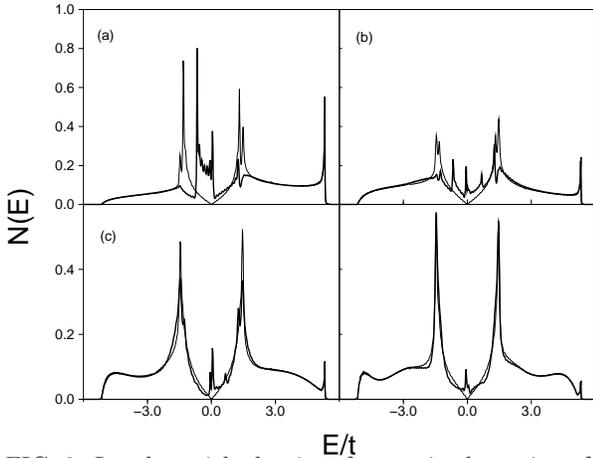} 
\caption{Local particle density of states in the region of the interface 
when the barrier height is weak, $V=t$. 
The plotted curves are otherwise exactly as in Fig. 8.\label{fignine}} 
\end{figure}

\begin{figure}
\includegraphics[scale=0.35,angle=-90]{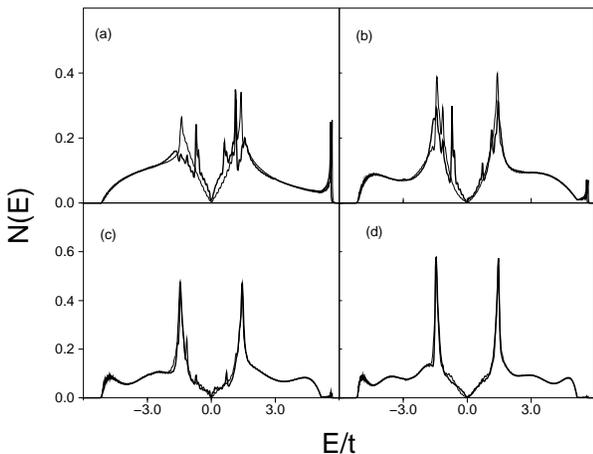} 
\caption{Local particle density of states in the region of the 
interface when the barrier height is  intermediate, $V=2t$. We 
consider both  a phase slip (thick solid line) and no phase slip 
(thin solid line). The curves plotted are (a) $x=1$, (b) $x=2$, 
(c) $x=3$ and (d) $x=4$.\label{figten}} 
\end{figure}

\begin{figure}
\includegraphics[scale=0.35,angle=-90]{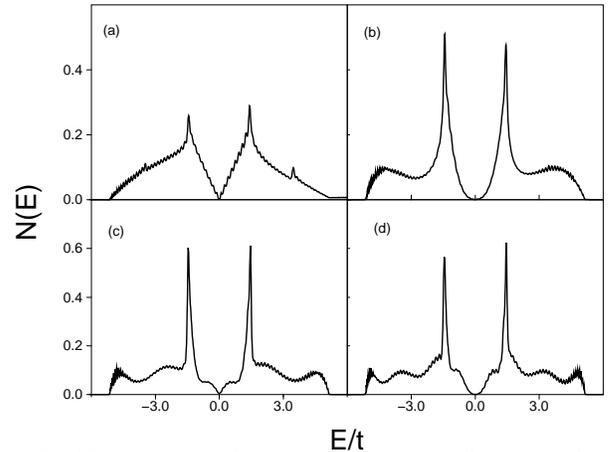} 
\caption{Local particle density of states in the region of the interface 
when the barrier height is strong, $V=5t$. 
The plotted curves are otherwise exactly as in Fig. 10.\label{figeleven}} 
\end{figure}

To see the physical origin of these three regimes we 
study the local particle density of states, $N(E)$, in the 
region of the interface. Figs.~\ref{figeight},~\ref{fignine},~\ref{figten} 
and ~\ref{figeleven} 
show the density of states 
$N(E)$ 
 for four strengths of barrier, 
$V/t=0$, $1$,  $2$ and  $5$, respectively. 
In these four figures we 
have plotted both the density of states in the 
presence of a phase slip, $N^\pi(E)$, (thick solid line) 
 and in the absence of a phase 
slip, $N^0(E)$, (thin solid line). We also show the densities of 
states for four values of $x$, the distance from the interface. 
In the first case we consider $V/t=0$ (Fig.~\ref{figeight}). The 
first thing to note is that the density of states in the absence 
of a phase slip corresponds to the pure bulk lattice $d$--wave 
density of states. This is simply because for $V=0$ and $\theta=0$ 
there is no interface. However the density of states in the 
presence of a phase slip, $N^{\pi}(E)$ does change in as we move 
away from the interface.  Comparing $N^0(E)$ and $N^{\pi}(E)$ we 
see a dramatic shift in the density of states in the region of the 
Fermi energy. We can see in Fig.~\ref{figeight}(a) that there is a 
strong shift of spectral weight from the peaks into the region 
near the Fermi level. We see that the gradient of the local 
particle density of states, in the region of $E=0$ is dramatically 
changed, and the size of the gap, (the distance between the two 
peaks) is strongly reduced in the presence of a phase slip. Moving 
away from $x=0$ (Figs 8(b),8(c) and 8(d)) we see that these large 
changes in density of states gradually diminish with $x$ until 
$N^0(E)$ and $N^{\pi}(E)$ become nearly identical. Interestingly 
the energy region near $E=0$ is most strongly affected when $x$ is 
large, corresponding to the long-range nature of the energy states 
associated with the $d$--wave gap node\cite{balatsky,leggett}. 

There is a very strong contrast between the case $V=0$ in 
Fig.~\ref{figeight} and the case $V/t=1$ shown in Fig.~\ref{fignine}. 
In  Fig.~\ref{fignine} we see that $N^0(E)$ (thin solid line) 
has no subgap structure at all, and this is 
quite different to the dramatic subgap 
structure obtained 
for $N^{\pi}(E)$.

As we increase the barrier strength to $V/t=2$ ( Fig.~\ref{figten}) we still 
see that there is no subgap structure in $N^0(E)$ whereas $N^{\pi}(E)$ 
still has some subgap structure, although 
this structure is not as dramatic as in 
 Fig.~\ref{fignine}. 
In  Fig.~\ref{figten} we have only plotted $N^0(E)$ 
and $N^{\pi}(E)$ for $x \neq 0$ 
because the $x=0$ density of states is dominated by  states well away 
from the gap region. 
 
In  Fig.~\ref{figeleven} we see that in the presence of a very 
strong barrier, $V/t=5$, $N^0(E)=N^{\pi}(E)$ suggesting that the 
two sides to the interface have become effectively decoupled. Also 
in this limit there is no resonant structure in either $N^0(E)$ or 
$N^{\pi}(E)$. 

Finally we estimate the Josephson critical current as a function 
of the barrier height, by determining the Josephson energy. This 
is defined by 
\begin{equation} 
 E_J = \frac{1}{2} \left(E(\pi) - E(0) \right) 
\end{equation} 
where we estimate the energy difference from the integrated 
densities of states 
 \begin{equation} 
 E(\phi) = \int_{-\infty}^0 N(\phi)(E) dE 
\end{equation} 
for $\phi=\pi$ and $\phi=0$ respectively. The critical current is 
approximately proportional to this energy difference, 
\begin{equation} 
I_c\propto E_J 
\end{equation} 
where the constant of proportionality depends on the form of the 
$E(\phi)$ relationship. In the simple Josephson tunneling case 
$E(\phi) - E(0) = E_J \sin{\phi} $, and $I_c=E_J$, but in general 
other forms of $E(\phi)$ are possible. For example we have found 
that $E(\phi)$ is approximately sawtooth in grain boundary 
junctions\cite{jason} and there is experimental evidence for 
non-sine wave behaviour\cite{illichev}. 
 
\begin{figure} 
\includegraphics[scale=0.35]{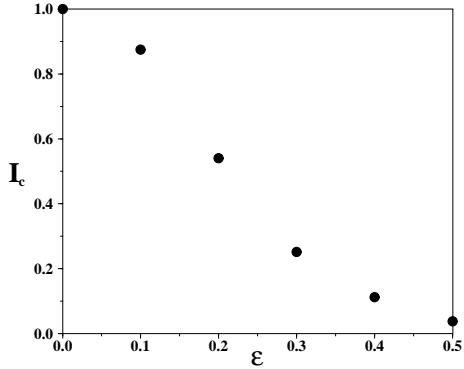} 
\caption{The normalized Josephson critical current 
as a function of interface barrier height, $V/t$.\label{figtwelve}} 
\end{figure}

In  Fig.~\ref{figtwelve} we have plotted the calculated critical current 
 ($I_c$) 
as a function of the strength of the impurity potential 
($\epsilon/t$) in region $I$. In this figure we have normalized 
the critical current with 
 respect to the critical current when no barrier is present. 
 What we again 
observe is consistent with the three different regimes (i), little 
or no normal scattering at the interface, where there is no sub 
gab resonant structure but $N^{\pi}(E) \ne N^0(E)$ for $|E| < 
|\Delta^{(d)}|$, (ii), moderate normal scattering at the 
interface, where there is sub gap resonant structure in 
$N^{\pi}(E)$ but not in $N^0(E)$ and finally (iii), large normal 
scattering at the interface, where $N^{\pi}(E) \approx N^0(E)$. 
From this it is possible to deduce that, as one would expect for a 
high strength interface barriers, the Josephson Critical Current 
approaches zero. The three regimes are visible 
in the calculated $I_c$ of Fig.~\ref{figtwelve}: 
 in the first weak decrease in $I_c$ 
with barrier height, the intermediate regime where there 
is a strong decrease in $I_c$, and in the strong scattering regime where 
$I_c \rightarrow 0$.

\section{Conclusions.} 
 
In this paper we have investigated the properties of a $d$--wave 
order parameter in the region of a (100) interface between two 
$d_{x^2-y^2}$ superconductors. We have determined how the barrier 
height at the interface influences the local density of states and 
Josephson current. In particular we could identify three distinct 
regimes with qualitatively different behaviour, corresponding to 
weak, intermediate and strong barriers, respectively. 
 
We have also calculated the subdominant extended $s$--wave, and 
$p$--wave like order parameter components at the interface.  In 
particular we have shown that the $p$--wave component has a 
temperature dependence which reveals the underlying bulk $T_c$ for 
$p$--wave pairing, even though this is normally hidden by the 
higher bulk $d$--wave $T_c$.  In principle a tunneling experiment 
could be carried out to test whether or not this $p$--wave 
component occurs in real systems. Such an experiment would provide 
qualitatively new information about the form of the microscopic 
pairing interaction.

\section{Acknowledgments.} 
 
This work was supported by the EPSRC under grant number GR/L22454 and the 
TMR network Dynamics of Nanostructures. 
We would like to thank B.L Gy\"orffy, J.J. Hogan O'Neill 
and J.P. Wallington for useful 
discussions.

\end{multicols} 

\begin{references}
\vspace{-1.5cm} 
\bibitem{Scal95}D.J. Scalapino, Phys. Rep. {\bf 250}, 329 (1995). 
\bibitem{Harl95}D.J. van Harlington, Rev. Mod. Phys. {\bf 67}, 515 (1995). 
\bibitem{jfa96} J.F. Annett, N. Goldenfeld, A.J. Leggett in Physical 
Properties of High Temperature Superconductors, Vol. 5, D.M. Ginsberg (ed.) 
(World Scientific, Singapore, 1996). 
\bibitem{kouznetsov} K.A. Kouznetsov et al., phys. Rev. Lett. {bf 79}, 
3050 (1997). 
\bibitem{goldman}A. Bhattacharya {\it et al.}, Phys. Rev. Lett. {\bf 82}, 
3132 (1999). 
\bibitem{MS1}M. Matsumoto and H. Shiba, J. Phys. Soc. Jpn. {\bf 64}, 3384 
(1995). 
\bibitem{MS2}M. Matsumoto and H. Shiba, J. Phys. Soc. Jpn. {\bf 64}, 4867 
(1995). 
\bibitem{MS3}M. Matsumoto and H. Shiba, J. Phys. Soc. Jpn. {\bf 65}, 2194 
(1996). 
\bibitem{us}A.M. Martin and J.F. Annett, Phys. Rev. B {\bf 57}, 8709 (1998). 
\bibitem{covington} M. Covington {\em et al.}, Phys. Rev. Lett. {\bf 79}, 
277 (1997). 
\bibitem{BBS}W. Belzig, C. Bruder and M. Sigrist, Phys. Rev. Lett. {\bf 80}, 
4285 (1998). 
\bibitem{YTanaka}Y. Tanaka, Phys. Rev. Lett. {\bf 72}, 3871 (1994). 
\bibitem{BGZ}Y.S Barash, A.V. Galaktionov and A.D. Zaikin, Phys. Rev. B 
{\bf 52}, 665 (1995). 
\bibitem{ZWT}J.X. Zhu, Z.D. Wang, and H.X. Tang, Phys. Rev. B {\bf 54}, 7354 
(1996). 
\bibitem{TWZ}H.X. Tang, Z.D. Wang and J.X. Zhu, Phys. Rev. B {\bf 54}, 12509 
(1996). 
\bibitem{Huck97}A. Huck, A. van Otterlo and M Sigrist, Phys. Rev. B {\bf 56}, 
14163 (1997). 
\bibitem{CRHu} C.R. Hu, Phys. Rev. Lett. {\bf 72}, 1526 (1994). 
\bibitem{XMT} J.H. Xu, J.H. Miller and C.S Ting, Phys. Rev. B {\bf 53}, 3604 
(1996). 
\bibitem{BSB}Y.S. Barash, A.A. Svidzinsky  and H. Burkhardt, Phys. Rev. B 
{\bf 55}, 15282 (1997). 
\bibitem{Purdue} A.M. Martin and J.F. Annett, Superlattices and 
Microstructures, {\bf 25} 1019 (1999). 
\bibitem{Tanuma1} Y. Tanuma, Y. Tanaka, M. Yamashiro and S. Kashiwaya, 
Physica C {\bf 282}, 1857 (1997). 
\bibitem{Tanuma2} Y. Tanuma, Y. Tanaka, M. Yamashiro and S. Kashiwaya, 
Physica C {\bf 293}, 234 (1997). 
\bibitem{interface}A. Gurevich and E.A. Pashitskii, Phys Rev. B {\bf 57}, 
13878 (1998). 
\bibitem{Jon1}J. F. Annett and J.P. Wallington, In {\it Symmetry and 
Pairing in Superconductors}, eds. M Asuloos and S. Kruchinin, NATO ASI 
Proceedings. Dordrecht (Kuluver, 1998). 
\bibitem{RHaydock}R. Haydock, in: Solid state Physics, {\bf 35}, 
eds. Ehrenreich, F. Seitz and D. Turnbull (Academic Press, New York, 1980). 
\bibitem{annett} J.F. Annett and N.D. Goldenfeld, J. Low Temp. Phys. 
{\bf 89}, 197 (1992). 
\bibitem{LMG}G. Litak, P. Miller and B.L. Gy\"orffy, Physica C {\bf 251}, 263 
(1995). 
\bibitem{jason}  J.J. Hogan-O'Neill, A.M. Martin and J.F. Annett, 
Phys. Rev. B {\bf 60}, 3568 (1999) 
\bibitem{yang} C.N. Yang, Phys. Rev. Lett. {\bf 63}, 2144 (1989). 
\bibitem{gyorffy}  B.L. Gy\"orffy, J. Staunton and M. Stocks, 
Phys. Rev. B {\bf 44}, 5190 (1991). 
\bibitem{annett} J.F. Annett, Adv. Phys. {\bf 39}, 83 (1990). 
\bibitem{OJ}O. Entin-Wohlman and J. Bar-Sagi, Phys. Rev. B {\bf 18}, 3174 
(1977). 
\bibitem{YM}Y. Tanaka and M. Tsukada, Phys. Rev. B {\bf 42}, 2066 (1990). 
\bibitem{needreference} F. Sols and J. Ferrer, Phys. Rev. B {\bf 49}, 
15913 (1994). 
\bibitem{balatsky} A.V. Balatsky, M.I. Salkola, Phys. Rev. Lett. 
{\bf 76}, 2386 (1996). 
\bibitem{leggett} I. Kosztin and A.J. Leggett, 
 Phys. Rev. Lett. {\bf 79},  135 (1997). 
\bibitem{illichev} E. Il'ichev {\it et al.}, Phys Rev Lett. {\bf 81}, 
894 (1998). 
\end{references}
\end{document}